\documentclass[structabstract]{aa}  
\usepackage{graphicx}
\usepackage{natbib}
\usepackage{txfonts}
\begin{document}
   \title{CV1 in the globular cluster M 22: confirming its nature through X-ray observations and optical spectroscopy}
\titlerunning{CV1 in M 22}

%   \subtitle{I. Overviewing the $\kappa$-mechanism}

   \author{N. A. Webb
          \inst{1,2}
          \and
          M. Servillat\inst{3,4}
          }

   \institute{Universit\'e de Toulouse;
    UPS-OMP; IRAP, Toulouse, France,\\
              \email{Natalie.Webb@irap.omp.eu}
         \and 
          CNRS; IRAP; 9 avenue du Colonel Roche, BP 44346,
  F-31028 Toulouse Cedex 4, France 
        \and
Laboratoire AIM (CEA/DSM/IRFU/SAp, CNRS, Universit\'e Paris 7 Denis Diderot), CEA Saclay, Bat. 709, 91191 Gif-sur-Yvette, France
         \and
          Harvard-Smithsonian Center for Astrophysics, 60 Garden Street, MS-67, Cambridge, MA 02138, USA
             }

   \date{}

% \abstract{}{}{}{}{} 
% 5 {} token are mandatory
 
  \abstract
  % context heading (optional)
  % {} leave it empty if necessary  
   {Observations of cataclysmic variables in globular clusters appear to show a dearth of outbursts compared to those observed in the field. A number of explanations have been proposed, including low mass-transfer rates and/or moderate magnetic fields implying higher mass white dwarfs than the average observed in the field.  Alternatively this apparent dearth may be simply a selection bias.  }
  % aims heading (mandatory)
   {We examine multi-wavelength data of a new cataclysmic variable, CV1, in the globular cluster M 22 to try to constrain its period and magnetic nature, with an aim at understanding whether globular cluster cataclysmic variables are intrinsically different from those observed in the field.}
  % methods heading (mandatory)
   {We use the sub-arcsecond resolution of the $Chandra$ ACIS-S to identify the X-ray counterpart to CV1 and analyse the X-ray spectrum to determine the spectral model that best describes this source.  We also examine the low resolution optical spectrum for emission lines typical of cataclysmic variables. Cross correlating the H$_\alpha$ line in each individual spectrum also allows us to search for orbital motion.}
  % results heading (mandatory)
   {The X-ray spectrum reveals a source best-fitted with a high-temperature bremsstrahlung model and an X-ray unabsorbed luminosity of 1.8$\times$10$^{32}$ erg s$^{-1}$ (0.3-8.0 keV), which are typical of cataclysmic variables.  Optical spectra reveal Balmer emission lines, which are indicative of an accretion disc.  Potential radial velocity in the H$_\alpha$ emission line is detected and a period for CV1 is proposed.}
  % conclusions heading (optional), leave it empty if necessary 
   {These observations support the CV identification.  The radial velocity measurements suggest  that CV1 
may have an orbital period of $\sim$7 hours, but further higher resolution optical spectroscopy of CV1 is needed to unequivocally establish the  nature of this CV and its orbital period.}   

   \maketitle
%
%________________________________________________________________

\section{Introduction}
\label{sec:intro}

Globular clusters (GCs) are very dense groups of
old stars, where the most massive stars of the cluster have already evolved away from the main sequence and into compact objects (e.g. neutron stars and white dwarfs).  In these dense environments, interactions between cluster members are frequent, and binary systems, including those containing compact objects, can form readily \citep[e.g.][]{hut92}.  Indeed the number  of neutron star X-ray binaries in Galactic GCs scales with the cluster encounter rate \citep{gend03b,pool03}, as do the millisecond pulsars \citep{abdo10}, which are believed to be the progeny of the neutron star X-ray binaries \citep{wijn98}.  Simulations  also show that a large number of close binaries (neutron star X-ray binaries and cataclysmic variables, CVs) should exist in GCs \citep[e.g.][]{dist94,davi97,ivan06} .

Close binaries  are extremely difficult to locate in the optical domain because of
over-crowding.  Two methods that can be exploited to detect close binaries are either (i) to observe the cluster in the X-ray domain where only about 100 sources can be expected \citep[e.g.][]{serv08a,serv08b,webb06} or (ii) to try to detect variability and/or optical outbursts from the systems \citep[e.g.][]{serv11,kalu10,kalu96}.  With these methods, dozens of interacting binaries in Galactic GCs have been discovered.

Thanks to the proximity of the white
dwarf and its companion in a CV, material is
accreted from the companion star and stored  in the accretion disc
around the white dwarf, whilst it loses enough angular momentum to
fall onto the compact object.  Outbursts are believed to occur when
too much material builds up in the disc, increasing both the density
and the temperature, until the hydrogen ionises and the viscosity
increases sufficiently for the material to fall onto the white dwarf
\citep[][etc]{osak74,meye81,bath81}.  Such outbursts are characterised by a steep rise in the flux by several orders
of magnitude.  Many types of field CVs show such outbursts every few weeks to years.  Comparing the small population of CVs identified in GCs to those known in the field has revealed a striking and unexplained difference. Only very few GC CV outbursts have been observed
\citep[e.g.][]{pare94,shara96,shara87}, and it is
unclear why this should be.

It was originally  suggested that GC CVs may be mainly
magnetic  \citep[e.g. the five magnetic CVs in][]{grin99}.  Magnetic CVs
have accretion discs that are either partially or totally disrupted by
the strong white dwarf magnetic fields, known as intermediate polars,
and polars respectively.  Material is channelled along the field lines
onto the white dwarf, although in the case of intermediate polars, a
truncated disc can exist, and these systems can undergo a limited
number of outbursts \citep[e.g.][]{nort89}.  However, more recently
it has been proposed that it may not simply be the magnetic field that
is responsible for the lack of outbursts. \cite{dobr06}
suggest that it may be due to a combination of low mass-transfer rates
($_\sim ^< 10^{14-15}$ g s$^{-1}$) and moderately strong white dwarf
magnetic moments ($_\sim ^> 10^{30}$ G cm$^{3}$) that could
stabilise the CV discs through truncation of inner regions \citep{meye94} and thus prevent most of them from
experiencing frequent outbursts.   \cite{ivan06} also suggests that the lack of outbursts are due to
higher white dwarf masses in GC CVs compared to those in
the field.  This would  imply that GC CVs are intrinsically different
from those in the field.  This could be due to the difference in the
formation mechanisms of GC and field CVs, where a
substantial fraction of cluster CVs are likely to be formed through
encounters, rather than from their primordial binaries \citep{ivan06}.  Alternatively, the observed phenomenon could simply be due to selection biases, where our knowledge of CVs is limited to a small population of the closest CVs that are frequently observed to outburst, as these are the easiest objects to detect.  It has recently been proposed that the majority of CVs may be short period, low mass-transferring systems that therefore show infrequent outbursts \cite[e.g.][]{uemu10}.  If this is the case, GC CVs may be no different from the field CV population \citep[e.g.][]{serv11}.

Sahu et al. (2001) presents an  event in the Galactic GC M~22 in which a star was observed to brighten by approximately
three magnitudes in the {\em Hubble Space Telescope (HST)} band F814W,
over about ten days, and to fade again on a similar timescale. Some
interspersed F606W images showed a similar brightening. \cite{sahu01}
argue that an unseen low-mass star in the cluster had lensed a
background bulge star. To test whether this was a lensing event,
\cite{ande03} used optical photometry to measure the proper motion of the star that had been
seen to brighten and determined that it was part of the cluster. From
its variability, its H$_\alpha$ emission, and its coincidence with the
{\it Einstein}/{\it Rosat} source X4/B, \cite{ande03} conclude that
this source, which they call CV1, was one of a very small number of
confirmed or probable dwarf nova eruptions seen in GCs
and the first to be found in such a low-concentration cluster.
\cite{bond05} confirm the variable nature of this source using a
four year light curve of the same source in \object{M 22}, based on an
analysis of accumulated data from the Micro-lensing Observations in
Astrophysics (MOA) survey.  From the regularity of the three outbursts detected between 1999 and 2004 and
their magnitude, \cite{bond05} also propose that this source may be a
cataclysmic variable.  From the X-ray properties and optical variability, \cite{hour11} propose that CV1 is a U Gem-type cataclysmic variable.

In this paper we present both {\em Chandra} and {\em XMM-Newton} archived data
for this cluster, along with low-resolution optical spectroscopy of CV1
and discuss the nature of this CV in the context of outbursts from
GC CVs.

\section{X-ray observations}
\label{sec:Xobs}

{\em Chandra} observations of the Galactic GC M~22 were made on 24 May 2005 for 16 ks using the ACIS-S3 cameras (ObsIDs: 7919 \& 8566).  We used CIAO version 3.4 software and the CALDB v3.4.0 set of calibration
files (gain maps, quantum efficiency, quantum efficiency uniformity,
effective area). We reprocessed the level-1 event files of
both observations without including the pixel randomisation that
is added during standard processing. This method slightly improves
the point-spread function (PSF).We removed cosmic-ray
events that could be detected as spurious faint sources using
the tool acis\_detect\_afterglow, and identified bad pixels with
the tool acis\_run\_hotpix. The event lists were then filtered for
grades, status, and good time intervals (GTIs), as in the standard processing. We selected events within the energy
range 0.3-10.0 keV.

To obtain a list of source candidates, we employed the
CIAO wavelet-based wavdetect tool for source detection in the
field of view covered by the four ACIS-I chips. The two energy
bands used were the 0.3-10.0 keV band, with all the events that allowed
us to detect the faintest sources, and 0.5-6.0 keV that had a
higher signal-to-noise ratio, so gave more significant detections.  We selected
scales of 1.0, 1.4, 2.0, 2.8, 4.0, and 5.6 pixels. The scales
were chosen to look for narrow PSF sources on-axis and ensure
optimal separation, and larger PSF sources at the edge of
the detectors, where the PSF is degraded. We selected a threshold
probability of 10$^6$, designed to give one false source per
10$^6$ pixels. This led to the detection of 50 source candidates.  We added to this 33 sources previously detected with {\em XMM-Newton} \citep[observation ID: 0112220201,][]{webb04} that fall inside the {\em Chandra} field of view.  We then used ACIS-Extract to determine the sources detected with a significance greater than 99.999\% \citep{broo10}.  We thus detected 39 sources, each with at least five counts.  These sources are given in Table~\ref{tab:XraySources}.

\begin{table}[!h]
\caption[{\em Chandra} X-ray sources in the direction of M 22]{Chandra X-ray sources in the direction of M 22 labelled from the centre of the cluster (1) outwards.  For X-ray sources detected using {\em XMM-Newton} the source number from \cite{webb04} is also given.  Right ascension, declination, and the 1 $\sigma$ positional error are provided, along with the counts detected (cnt) and the source flux from ACIS-Extract (in $\mathrm{~erg~cm^{-2}~s^{-1}}$)}
\label{tab:XraySources}
\centering
\vspace{0.2cm}
\begin{tabular}{@{~}c@{~~}c@{~}|@{~}c@{~~}c@{~~}c@{~~}c@{~~}c@{~~}}
\hline
\hline
Src & Src & R.A. & Dec. & error & cnt & flux \\
Ch. & XMM & $^h$ $^m$ $^s$ &  $^\circ$ $\arcmin$ $\arcsec$ &  $\arcsec$      &     &   ($\times$10$^{-14}$)     \\
\hline
   $  1$ & -- & $ 18^h36^m24.21^s$ & $-23\degr54\arcmin10.13\arcsec$ & $ 
0.45$ & $   10$ & $     0.79 \pm     0.70$ \\
   $  2$ & 36 & $ 18^h36^m24.71^s$ & $-23\degr54\arcmin35.61\arcsec$ & $ 
0.41$ & $  231$ & $    18.87 \pm     3.50$ \\
   $  3$ & -- & $ 18^h36^m25.44^s$ & $-23\degr54\arcmin51.62\arcsec$ & $ 
0.48$ & $    7$ & $     0.52 \pm     0.17$ \\
   $  4$ & -- & $ 18^h36^m27.20^s$ & $-23\degr54\arcmin25.97\arcsec$ & $ 
0.43$ & $   27$ & $     2.13 \pm     1.34$  \\
   $  5$ & 33 & $ 18^h36^m21.70^s$ & $-23\degr53\arcmin34.99\arcsec$ & $ 
0.48$ & $    5$ & $     0.28 \pm     0.18$ \\
   $  6$ & 39 & $ 18^h36^m24.86^s$ & $-23\degr55\arcmin14.92\arcsec$ & $ 
0.42$ & $   56$ & $     5.00 \pm     2.08$ \\
   $  7$ & -- & $ 18^h36^m28.26^s$ & $-23\degr55\arcmin35.25\arcsec$ & $ 
0.48$ & $   11$ & $     0.66 \pm     0.47$ \\
   $  8$ &--  & $ 18^h36^m32.88^s$ & $-23\degr53\arcmin05.35\arcsec$ & $ 
0.44$ & $   19$ & $     1.12 \pm     0.71$  \\
   $  9$ & 32 & $ 18^h36^m18.40^s$ & $-23\degr52\arcmin17.19\arcsec$ & $ 
0.42$ & $   35$ & $     2.75 \pm     1.31$  \\
   $ 10$ & 23 & $ 18^h36^m28.29^s$ & $-23\degr56\arcmin24.64\arcsec$ & $ 
0.45$ & $   29$ & $     2.48 \pm     1.33$  \\
   $ 11$ & -- & $ 18^h36^m13.84^s$ & $-23\degr54\arcmin45.46\arcsec$ & $ 
0.53$ & $    6$ & $     0.39 \pm     0.30$  \\
   $ 12$ & -- & $ 18^h36^m22.33^s$ & $-23\degr51\arcmin40.08\arcsec$ & $ 
0.46$ & $    8$ & $     0.89 \pm     0.84$  \\
   $ 13$ & 55 & $ 18^h36^m14.44^s$ & $-23\degr55\arcmin38.86\arcsec$ & $ 
0.46$ & $   24$ & $     1.87 \pm     1.01$  \\
   $ 14$ & -- & $ 18^h36^m14.37^s$ & $-23\degr52\arcmin41.76\arcsec$ & $ 
0.48$ & $    8$ & $     1.01 \pm     0.75$  \\
   $ 15$ & -- & $ 18^h36^m24.10^s$ & $-23\degr57\arcmin11.20\arcsec$ & $ 
0.66$ & $    7$ & $     0.37 \pm     0.22$  \\
   $ 16$ & 37 & $ 18^h36^m38.38^s$ & $-23\degr54\arcmin47.42\arcsec$ & $ 
0.46$ & $   23$ & $     1.86 \pm     1.16$  \\
   $ 17$ & -- & $ 18^h36^m33.80^s$ & $-23\degr51\arcmin37.50\arcsec$ & $ 
0.45$ & $   14$ & $     0.80 \pm     0.38$  \\
   $ 18$ & 34 & $ 18^h36^m39.06^s$ & $-23\degr53\arcmin46.63\arcsec$ & $ 
0.44$ & $   32$ & $     1.93 \pm     0.51$  \\
   $ 19$ & -- & $ 18^h36^m12.17^s$ & $-23\degr56\arcmin23.69\arcsec$ & $ 
0.64$ & $    7$ & $     0.39 \pm     0.30$  \\
   $ 20$ & 20 & $ 18^h36^m15.44^s$ & $-23\degr50\arcmin56.80\arcsec$ & $ 
0.48$ & $    9$ & $     1.40 \pm     0.99$  \\
   $ 21$ & -- & $ 18^h36^m30.14^s$ & $-23\degr49\arcmin58.78\arcsec$ & $ 
0.53$ & $    6$ & $     0.66 \pm     0.26$  \\
   $ 22$ & 73 & $ 18^h36^m21.04^s$ & $-23\degr49\arcmin19.50\arcsec$ & $ 
0.56$ & $    6$ & $     0.48 \pm     0.33$  \\
   $ 23$ & 71 & $ 18^h36^m25.50^s$ & $-23\degr59\arcmin09.31\arcsec$ & $ 
0.75$ & $   16$ & $     1.15 \pm     0.82$  \\
   $ 24$ & -- & $ 18^h36^m42.94^s$ & $-23\degr56\arcmin58.94\arcsec$ & $ 
0.61$ & $   18$ & $     1.67 \pm     1.03$  \\
   $ 25$ &  8 & $ 18^h36^m34.63^s$ & $-23\degr49\arcmin29.11\arcsec$ & $ 
0.46$ & $   25$ & $     2.21 \pm     1.29$  \\
   $ 26$ & 19 & $ 18^h36^m 5.64^s$ & $-23\degr50\arcmin50.27\arcsec$ & $ 
0.62$ & $    8$ & $     2.23 \pm     2.03$  \\
   $ 27$ & 44 & $ 18^h36^m18.31^s$ & $-24\degr00\arcmin55.88\arcsec$ & $ 
0.70$ & $   48$ & $     4.98 \pm     2.20$  \\
   $ 28$ & 16 & $ 18^h36^m15.15^s$ & $-23\degr46\arcmin13.70\arcsec$ & $ 
0.69$ & $   21$ & $     2.18 \pm     1.04$  \\
   $ 29$ & 17 & $ 18^h36^m10.45^s$ & $-23\degr46\arcmin27.38\arcsec$ & $ 
0.63$ & $   30$ & $     3.24 \pm     0.79$  \\
   $ 30$ & -- & $ 18^h37^m 0.63^s$ & $-23\degr56\arcmin12.81\arcsec$ & $ 
0.93$ & $   28$ & $     2.44 \pm     1.23$  \\
   $ 31$ & -- & $ 18^h36^m59.94^s$ & $-23\degr51\arcmin29.11\arcsec$ & $ 
0.74$ & $   39$ & $     4.76 \pm     2.74$  \\
   $ 32$ & -- & $ 18^h36^m56.04^s$ & $-23\degr48\arcmin44.82\arcsec$ & $ 
1.06$ & $   14$ & $     1.75 \pm     0.77$  \\
   $ 33$ & 29 & $ 18^h36^m27.68^s$ & $-23\degr45\arcmin06.61\arcsec$ & $ 
0.57$ & $   64$ & $     6.46 \pm     2.26$  \\
   $ 34$ & 25 & $ 18^h37^m 0.56^s$ & $-23\degr58\arcmin29.70\arcsec$ & $ 
1.28$ & $   21$ & $     1.93 \pm     1.12$  \\
   $ 35$ & -- & $ 18^h36^m56.82^s$ & $-23\degr48\arcmin08.59\arcsec$ & $ 
1.25$ & $   12$ & $     1.23 \pm     1.12$  \\
   $ 36$ & -- & $ 18^h36^m30.56^s$ & $-23\degr44\arcmin27.38\arcsec$ & $ 
1.03$ & $   21$ & $     1.81 \pm     1.11$  \\
   $ 37$ & -- & $ 18^h36^m56.20^s$ & $-23\degr47\arcmin27.39\arcsec$ & $ 
1.49$ & $   10$ & $     0.70 \pm     0.48$  \\
   $ 38$ & 43 & $ 18^h36^m59.48^s$ & $-24\degr00\arcmin28.59\arcsec$ & $ 
1.23$ & $   34$ & $     5.14 \pm     1.68$  \\
   $ 39$ & -- & $ 18^h37^m 8.63^s$ & $-23\degr50\arcmin27.36\arcsec$ & $ 
1.15$ & $   45$ & $     4.50 \pm     2.28$ \\
\hline
\end{tabular}
\end{table}

\begin{figure}
  \centering\includegraphics[angle=0,width=9.0cm]{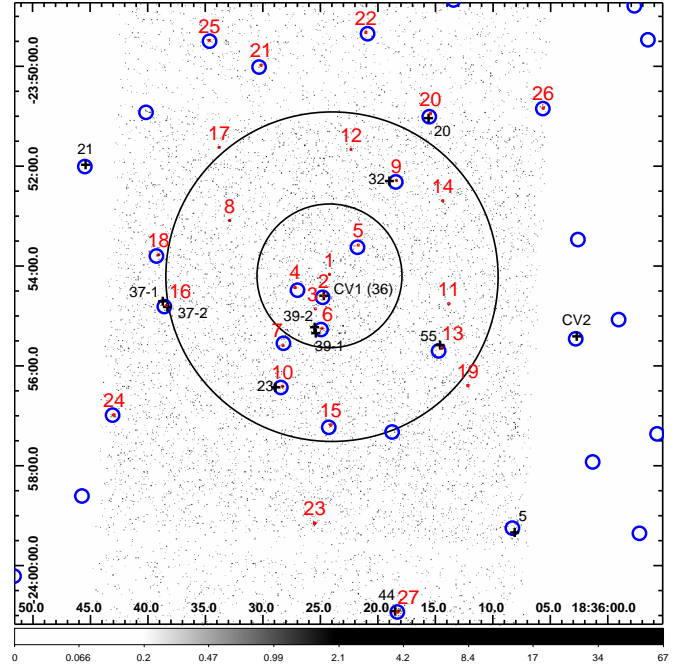}
\caption{X-ray sources in the direction of M~22.  Chandra sources, with the numbers given in Table~\ref{tab:XraySources} are shown with their 1 $\sigma$ error circles (red).  The larger (blue) error circles show the {\em XMM-Newton} sources detected in the field of view, as extracted from the catalogue 2XMM \citep{wats09}.  The crosses show the positions of the possible optical counterparts.  The two largest circles centered on the image are the core radius (1.33\arcmin) and the half mass radius (3.36\arcmin) \citep[][December 2010 revision]{harr96}.}
\label{fig:XMMChandra}
\end{figure}

Using the n$_H$ value of 2.2$\times$10$^{21}$ cm$^{-2}$ \citep{john94} and a power law slope of 2.1 (the average of the 39 detected sources), the unabsorbed flux limit of the {\em Chandra} observation is  3 $\times$ 10$^{-15}$ $\mathrm{~erg~cm^{-2}~s^{-1}}$ (0.5-8.0 keV).  Using the distance measurement of 3.2 kpc \citep[][December 2010 revision]{harr96} this translates to an unabsorbed luminosity limit for an object in the cluster of 4 $\times$ 10$^{30}$ $\mathrm{~erg~s^{-1}}$ (0.5-8.0 keV).  This is slightly deeper than the {\em XMM-Newton} observation presented in \cite{webb04}, which has an unabsorbed luminosity limit in the {\em Chandra} band and using a distance of 3.2 kpc (0.6 kpc greater than the distance given in \cite{webb04}) of 7 $\times$ 10$^{30}$ $\mathrm{~erg~s^{-1}}$.  As a result, more sources are detected within the  {\em Chandra} field of view.  All but two of the {\em XMM-Newton} sources that fall within the {\em Chandra} field of view were detected, see Fig.~\ref{fig:XMMChandra}. These two sources, which are below the threshold presented in \cite{webb04} but present in the more recently analysed data of the {\em 2XMM} catalogue \citep{wats09}, may be variable sources.  We find that Chandra source 3 is coincident with the position of a newly discovered binary radio pulsar, M22A \citep{lync11}, to within the 2 $\sigma$ error radius of the X-ray source. None of the sources
correspond to the two new radio sources that are proposed as black hole binaries \cite{stra12}, but the X-ray and radio
observations are not simultaneous, which may account for their not being detected.

Thanks to the superior angular resolution of {\em Chandra}, the 1 $\sigma$ positional error of the {\em Chandra} sources  is, for the majority of the sources, sub-arcsecond (see Table~\ref{tab:XraySources}), compared to the 90\% confidence mean statistical error of $\langle$ 7.6$\arcsec\rangle$ for the {\em XMM-Newton} sources \citep{webb04}.  This difference is clearly seen in Fig.~\ref{fig:XMMChandra}, where both the {\em Chandra} and the {\em XMM-Newton} sources are presented, along with their error circles.  Also included in  Fig.~\ref{fig:XMMChandra} are the possible optical counterparts selected following a search for sources detected with the {\em Anglo Australian Telescope} ({\em AAT}) {\em Wide Field Imager} that were {\em bluer} than the GC main sequence stars and that fall within the {\em XMM-Newton} X-ray error circles \citep{webb04} and CV1. It is evident from this figure that the majority of the possible optical counterparts are unlikely to be the actual counterpart, as they are not within the smaller {\em Chandra} error circle.  The position of the source CV1 falls at 18$^h$36$^m$24$\fs$70, -23$^\circ$54$\arcmin$35$\farcs$1, with a 1 $\sigma$ error circle of $\sim$1$\arcsec$ \citep{bond05}, is, however, coincident within the errors with the {\em Chandra} source, see Fig.~\ref{fig:XMMChandra}.

\begin{figure}
  \centering\includegraphics[angle=-90,width=9.0cm]{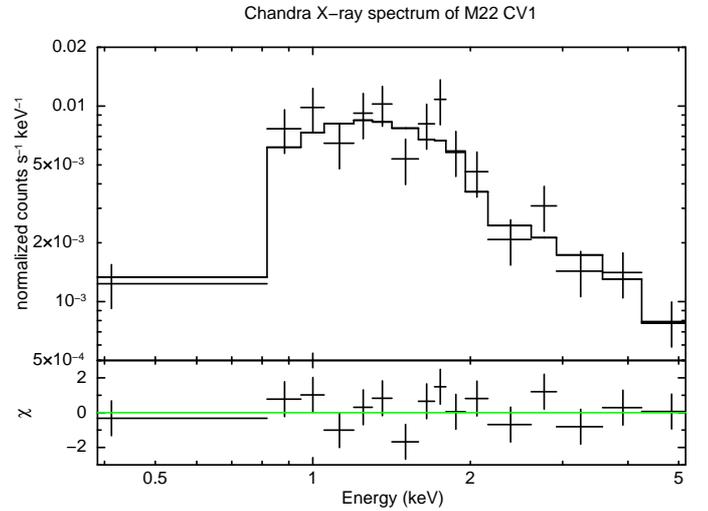}
\caption{Upper: {\em Chandra} spectrum of CV1, fitted with a bremsstrahlung model, as described in Sec.~\ref{sec:Xobs}.  Lower: Residuals to the fit.}
\label{fig:Xrayspec}
\end{figure}

To confirm that {\em Chandra} source 2 is the X-ray counterpart, we extracted the X-ray spectrum, and fitted this with simple models.  Equally good fits are obtained for a power law fit (n$_H$=1.95($\pm^{\scriptscriptstyle 4.73}_{\scriptscriptstyle 0.30}$)$\times$10$^{21}$ cm$^{-2}$, $\Gamma$=1.47$\pm^{\scriptscriptstyle 0.42}_{\scriptscriptstyle 0.44}$, $\chi^{\scriptscriptstyle 2}_{\scriptscriptstyle \nu}$=1.02, 12 dof, errors are 90\%) or a bremsstrahlung (n$_H$=1.62($\pm^{\scriptscriptstyle 1.90}_{\scriptscriptstyle 1.62}$)$\times$10$^{21}$ cm$^{-2}$, kT=15.66$\pm^{\scriptscriptstyle 94.0}_{\scriptscriptstyle 9.73}$ keV, $\chi^{\scriptscriptstyle 2}_{\scriptscriptstyle \nu}$=0.99, 12 dof). These fits are consistent with those found by \cite{hour11} with the same data sets. Using the latter fit we find an unabsorbed flux of 1.49($\pm^{\scriptscriptstyle 0.25}_{\scriptscriptstyle 0.45}$)$\times$10$^{-13}$ erg cm$^{-2}$ s$^{-1}$ (0.3-8.0 keV, error 1 $\sigma$), which translates to an unabsorbed luminosity of  1.83$\times$10$^{32}$ erg  s$^{-1}$ (0.3-8.0 keV) at the distance of the cluster.  The data and the bremsstrahlung fit can be seen in Fig.~\ref{fig:Xrayspec}.

\section{Optical observations and data reduction}
\label{sec:obs}

Observations were made with the VIsible Multi-Object Spectrograph
(VIMOS) on the 8.2 m Melipal telescope of the {\em European Southern
Observatory Very Large Telescope} ({\em VLT}), with the aim of identifying the X-ray sources given in \cite{webb04}.  For further information about
the instrument used here see \citet{lefe04}.  The pre-images were taken on
14 April 2004 (field 1, centred on 18:35:58.2, -23:57:39.9, J2000)
and on 16 May 2004 (field 2, centred on 18:36:34.0, -23:57:39.9,
J2000).  These images were made using the U filter and were of 120
and 60 seconds respectively.  The positions from the {\em AAT} data were transformed
onto the VIMOS instrument coordinate system using the {\em VLT} supplied
{\em VIMOS Mask Preparation Software} (VMMPS) and then masks for the spectroscopic observations were defined with the same package.  Four
fairly bright stars that were well spaced out on each of the four CCDs were
allotted slits to be used for alignment purposes.  We placed slits of
1\arcsec\ width (4.88 pixels) and 14\arcsec\ in length (68 pixels) on each
of the targets.  The remaining slits were placed on random other stars in
the field of view.

The spectroscopic observations were carried out in service mode on
three different nights, June 9 and 22 and July 9 2004, and are
summarised in Table~\ref{tab:obs}. The blue grism has a  range of
3700-6700\AA, and  the ESO VIMOS web pages give a spectral resolution (${\mathrm{\lambda/\Delta\ \lambda}}$) of 180 for a 1\arcsec\ slit, with a
dispersion of 5.35 \AA/pixel.  The red grism  has a  range of
5500-9500\AA, and the ESO VIMOS web pages give a spectral resolution of 210 for a 1\arcsec\ slit and a
dispersion of 7.14 \AA/pixel.  The seeing was typically below
0.8\arcsec, except for the last night when it exceeded 1\arcsec.
Flat-field lamp and helium-argon arc lamp observations were taken to
flat-field and wavelength-calibrate the data. The flux standard stars
\object{LTT 1020}, \object{NGC 7293}, and \object{Feige 110} were also
observed during the night to flux calibrate the targets.

\begin{table}[!h]
  \caption[]{Summary of the optical observations presented}
     \label{tab:obs}
       \begin{tabular}{lccc}
         \hline 
          \hline
          \noalign{\smallskip}  Date     & Field & Grism  &
         Observations \\  \hline   June 9 2004 & 2 & red & 650,
         350, 4$\times$150, 30 secs \\ June 22 2004 & 1 & blue &
         4$\times$650 \\ June 22 2004 & 2 & blue & 4$\times$650 \\
         July 9 2004 & 1 & red & 4$\times$650 \\ \hline
  \end{tabular}
 \end{table}

The data were reduced using VIPGI (VIMOS Interactive Pipeline and
Graphical Interface, \citealt{scod05}).  The data were bias-subtracted,
but the dark current was negligible ($<$1 e$^{-}$/pixel), so no dark
correction was carried out.  Flat-field corrections were also made and
the spectra were wavelength-calibrated using the arc lamps, which gave
the wavelength measurements good to approximately 1\AA.    Due to the
very high density of sources, each spectrum was extracted using an
IDL program written specifically to estimate the sky contribution in
each slit and then to use an optimal extraction method \citep{horn86}
to obtain the spectrum.  The flux standards that had been observed
throughout the observations, on each CCD individually and in both
wavelength regions were extracted in a similar manner.  Using these
observations, coupled with the spectral energy distributions given in
\cite{oke90} and \cite{hamu94}, we corrected for the instrument
response and flux-calibrated the spectra.  We also dereddened the
spectra using  E(B-V)=0.38$\pm$0.02 \citep{mona04}.

In the following we concentrate on the optical spectra of CV1, as this counterpart is the only one that falls within the {\em Chandra} error circle, see Section~\ref{sec:Xobs}.  These spectra were taken during the field 2 observations, see Table~\ref{tab:obs}.

\section{The candidate cataclysmic variable CV1}

\begin{figure}
  \centering\includegraphics[angle=-90,width=9.2cm]{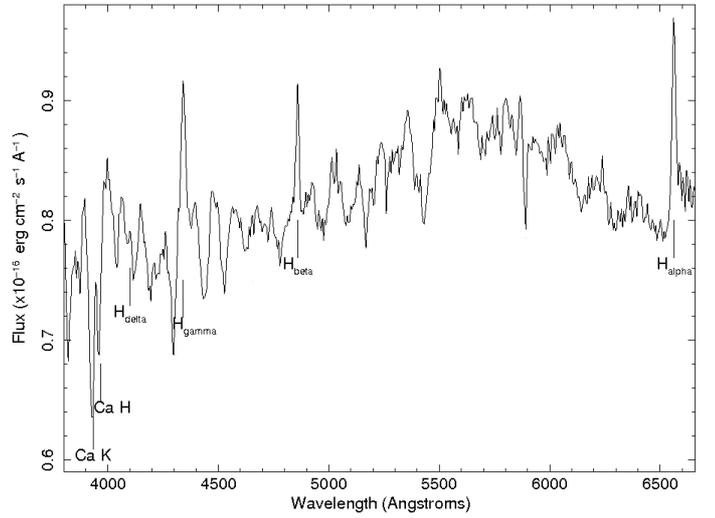}
\caption{Dereddened blue spectrum of the optical counterpart to source 36.  The strongest Balmer lines are seen in emission.  Also evident are the Ca H and K lines (rest wavelengths 3968.5\AA\ and 3933.7\AA\ respectively).}
\label{fig:src36spec}
\end{figure}                

\begin{figure}
  \centering\includegraphics[angle=-90,width=9.2cm]{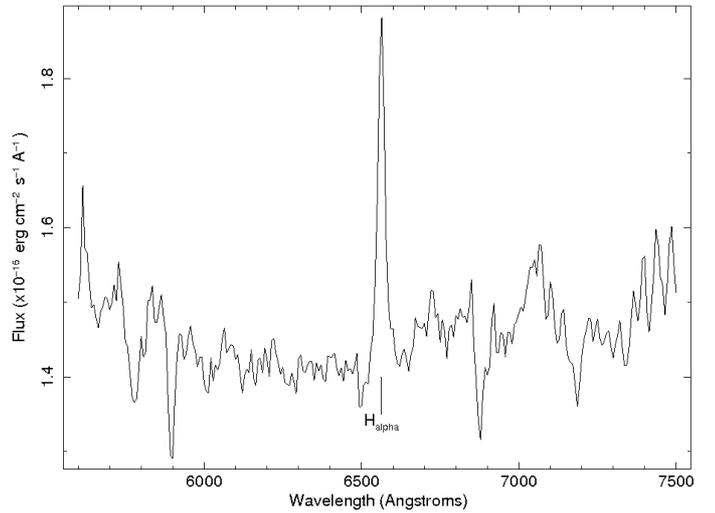}
\caption{Dereddened red spectrum of the optical counterpart to source 36.  The H$_\alpha$ line as well as a possible He I (5876\AA) are seen in emission. Only the spectrum up to 7500\AA\ is presented as redder than this the telluric absorption and fringing makes the spectrum very noisy.}
\label{fig:src36specred}
\end{figure}                

We extracted  optical spectra of CV1 as described in Section~\ref{sec:obs}.  We estimated that
the brighter neighbour contributes approximately 10\% to our source's
spectrum, by plotting the point spread function of the two sources relative to each other and measuring the amount of flux from the neighbour within the point spread function of CV1. We then subtracted this flux from our target. The
neighbouring star's spectrum contains no emission lines and appears to
be a mid K-type star. To check that this procedure introduced no strange features into our spectra, we extracted the CV1 spectra using only the region in the slit where CV1 dominated the emission.  We therefore did not need to subtract the close neighbour. This obviously has an impact on the total flux. With this method we extracted spectra with only a third of the flux, but the spectra using the two methods are none the less very similar. The optical spectrum of the target (see
Fig.~\ref{fig:src36spec} and Fig.~\ref{fig:src36specred}) shows Balmer line emission that are indicative of accretion.  We
give the equivalent widths of the principal Balmer lines (\AA) in
Table~\ref{tab:eqwidths} along with their fluxes. We have identified the Ca II H and
K absorption lines at 3968.5 and 3933.7 \AA, which we assume are due
to the late type companion star (see below).  There may also be some evidence for weak Ti0 bands in the red spectrum at 7055\AA, which would indicate that it is an early M-star if they come from the secondary. We also note that the flux in the red spectrum is slightly higher than that presented in the blue spectra. CV1 was declining from an outburst when the red spectra were taken \citep{hour11}, whereas it was in quiescence when the blue spectra were taken \citep{hour11}, which may explain the variation in flux.

\begin{table}[!h]
  \caption[]{Equivalent widths of the principal spectral lines in the optical spectrum of CV1. The errors are of the order of 20\%.}
     \label{tab:eqwidths}
       \begin{tabular}{ccc}
         \hline 
          \hline
           \noalign{\smallskip}  Line    & Eq. Width  & Flux \\ &
         (\AA) & ($\times$10$^{-16}$ erg cm$^{-2}$ s$^{-1}$
         \AA$^{-1}$)\\  \hline     H$\alpha$ & -5.6 & 3.8
         \\ H$\beta$ & -2.1 & 1.6 \\ H$\gamma$ & -3.7 & 2.8
         \\ H$\delta$ & -1.1 & 0.8 \\ 
%He I (5768\AA)& -0.5 & 1.0 \\ He
%         II (4686\AA) & -1.4 & 3.9 \\ 
\hline
  \end{tabular}
 \end{table}

Seven observations were made of field 2 with the red grism  and four for field 1. Four observations were also made of each of the fields observed with the blue grism, see Table~\ref{tab:obs}. Multiple, short observations were made so as not to
saturate the brightest counterparts with a single long exposure.  We have taken advantage of the
fact that the individual optical spectra of CV1 contain enough
counts to identify the main emission lines in all spectra.  We first subtracted
the continuum before cross-correlating the region around the H$_\alpha$
line (6540-6600\AA, as this region is present in both the red and the
blue spectra), with the total spectrum (summed red spectra)  of a late G/early K-type star believed to be in the cluster, using the {\em IRAF} \citep{tody93} task {\em fxcor}\footnote{http://iraf.noao.edu/}.   The results are presented in
Fig.~\ref{fig:src36rv} with the errors bars given using the {\em fxcor}. A critical analysis of these errors is given below.  We show a fit that assumes  a circular orbit with an amplitude  of 35.96$\pm^{\scriptscriptstyle 8.71}_{\scriptscriptstyle 16.00}$ km s$^{-1}$ and a period of 7.44$\pm^{\scriptscriptstyle 0.03}_{\scriptscriptstyle 0.19}$ hours ($\chi^{\scriptscriptstyle 2}_{\scriptscriptstyle \nu}$=1.2  (8 dof), errors are given at the 90\% level).  Other periods between $\sim$2.5 and 24 hours give acceptable fits i.e. $\chi^{\scriptscriptstyle 2}_{\scriptscriptstyle \nu} \simeq$1.2-2.0  (8 dof), indicating that, although the error on the period is small, there are other periods that can give almost as good a fit and therefore may be equally valid.  Cross correlating with another star in the cluster means that we can not measure the proper motion of the cluster itself. These observations suggest that we may have detected
orbital motion.  Although less significant, we see a similar trend from the other Balmer
lines visible in the blue spectra, which are likely to be coming from the disc and which appear to be an opposite trend
from the calcium H- and K-lines, as expected if these lines originate
in the secondary star.  

To investigate the reliability of the {\em fxcor} errors, we also estimated errors by considering the accuracy with which the central wavelength of the H$_\alpha$ line can be estimated. We fitted a Gaussian to the H$_\alpha$ line and determined the 90\% error on the the central wavelength. We then used that to calculate the error on the cross correlation. This error is much larger and varies between 18 and 90 km s$^{-1}$.  Fitting the radial velocity data with these larger error bars with either a constant (i.e. assuming no variability expected), gives a $\chi^{\scriptscriptstyle 2}_{\scriptscriptstyle \nu}$=0.14 (10 dof). Fitting with a constant+sinusoid (i.e. assuming some radial velocity variability) gives $\chi^{\scriptscriptstyle 2}_{\scriptscriptstyle \nu}$=0.03 (8 dof). This suggests that constant+sinusoid is preferred, but the reduced chi-squared values are very low, due to the inordinately large errors.  Fitting the data with the errors given by {\em fxcor} we find  $\chi^{\scriptscriptstyle 2}_{\scriptscriptstyle \nu}$=9.70  (10 dof) for a simple constant and $\chi^{\scriptscriptstyle 2}_{\scriptscriptstyle \nu}$=1.2  (8 dof) for a constant + sinusoid.  Again, the sinusoid is preferred.  We checked the {\em fxcor} cross correlation procedure for our data by cross correlating the summed late G-/early K-star with individual spectra of the same source (in both the blue and the red) and find that the position of the H$\alpha$ line does not vary to within $\pm$8 km s$^{-1}$. Cross correlating the same star with CV1 reveals radial velocities between -61 and +17 km/s (see Fig.~\ref{fig:src36rv}). It therefore seems probable that we have detected radial velocity variability in CV1 and that the errors could be considered to be as low as $\pm$8 km s$^{-1}$, which is  similar to the {\em fxcor} errors.

\begin{figure}
   \centering \includegraphics[angle=-90,width=8.0cm]{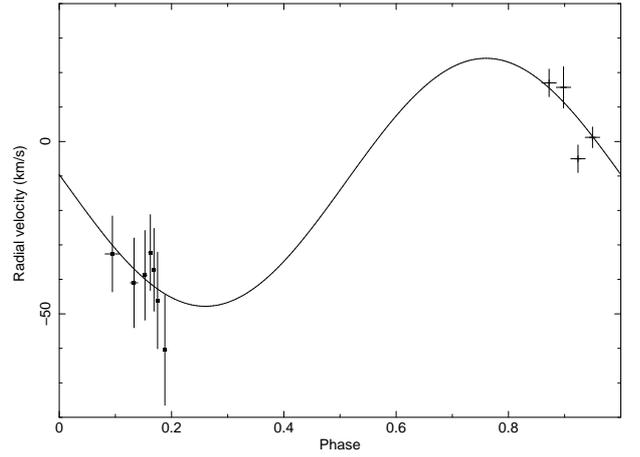}
\caption{Radial velocities and errors of the H$_\alpha$ line in the optical spectrum of CV1.  The solid line shows the best sinusoidal fit to these data. The points shown with filled squares are the radial velocity measurements made with the red spectra. The remaining four data points are the radial velocity measurements made with the blue spectra.}
\label{fig:src36rv}
\end{figure}

\section{Discussion and conclusion}

The {\em Chandra} source 2 and CV1  are coincident within the errors.
Employing the method used in \cite{zane08}, we calculated the
probability that CV1 is aligned by chance coincidence with {\em
  Chandra} source 2.  This probability is given as 1-e$^{-\pi \mu
  r^2}$, where $\mu$ is the measured object density in the {\em
  Chandra} field of view, and $r$ is the radius of the {\em HST} error
circle.  We find a probability of a chance coincidence of 2.34$\times$
10$^{-4}$, confirming CV1 as the optical counterpart to source 2.  The
X-ray flux and spectrum of source 2 are consistent with that of known
CVs \citep[e.g.][]{bask05,byck10} and supports the hypothesis that the X-ray source is a CV and therefore the X-ray counterpart to CV1. The
dereddened optical spectrum presented here substantiate that CV1 exhibits a blue excess compared to a main sequence star,
shows Balmer emission lines endorsing the CV nature.  \cite{ande03} note that the object they observed using photometry was redder than the main sequence, however, this may be due in part to the H$_\alpha$ emission, which is present in the V$_{606}$, R$_{675}$, and I$_{814}$ broad band filters they used.  We also see that the emission is relatively strong in the red.

From the magnitudes presented in \cite{ande03} and \cite{hour11} we can determine the expected period as follows. The absolute magnitude of this source using the distance of
\citet[][December 2010 revision]{harr96} is around 6.3 (V-band and
assuming that the CV is in M~22).  Using the correlation between
brightness and orbital period \citep{warn87}, in the same way as
\cite{ande03}, but using the revised distance and therefore revised
absolute magnitude, implies a period of around 10-11 hours. Taking the
peak V-magnitude of around three magnitudes brighter than the quiescent
value \citep{ande03} indicates a similar period.  However,
\cite{patt11} presents a revised version of this correlation, by using
precise distances to the 46 CVs studied.  He then extrapolates this to
study the shortest period CVs.  Without correcting CV1 for its
inclination, which is at present unknown, the \cite{patt11}
relationship reveals a period of around nine hours. The data presented in this paper could also be fitted with a period of around nine hours, however, the $\chi^{\scriptscriptstyle 2}_{\scriptscriptstyle \nu}$ then rises from 1.20 to 1.91 (8 dof) for the fit with a longer period.  However, there is a
lot of scatter in the data points presented in \cite{patt11}, so a
peak outburst magnitude of M$_V$=3.2 is still compatible with the
period of about seven hours.  It should also be noted that \cite{ande03}
show that CV1 is redder than the main sequence, as mentioned above.  They state that this may indicate
a secondary that could be larger than a normal main-sequence star, which may raise the question of the validity of these relations for this particular source and indeed we do see that this CV is fairly red.  More and higher resolution, higher signal-to-noise observations will be
necessary to confirm the true nature of this CV and validate its
period. 

As outlined in Section~\ref{sec:intro}, it has been proposed on numerous
occasions that GC CVs may be magnetic
in nature, which may help explain the dearth of outbursts observed from
these objects.     A period of around seven
hours may also lend further support for this type of object.  \cite{dobr06} propose that the
dearth of outbursts in GC CVs may be due to both low
mass-transfer rates and truncated inner discs, as seen in IPs. CV1 appears to show fewer outbursts than many of the well studied CVs, i.e. \cite{shar96b}, which presents 21 well studied CVs that have an average of $\sim$29 days between outbursts and an average duration of $\sim$15 days.  It would then be expected to observe about eight outbursts per year from these CVs on average.  Comparing the three outbursts that were observed between 1999 and 2004, if they were indeed the only outbursts during this time, CV1 appears to rarely show outbursts. However, the CVs presented in \cite{shar96b} have been well studied because they  show regular outbursts. It is likely that the majority of CVs show quite distinct characteristics, with much less regular outbursts  \citep{shar05}.  This would then imply  that the lack of observed GC CV outbursts is simply related to selection biases.

\begin{acknowledgements}
The authors are extremely grateful to the anonymous referee who raised a number of very valuable points and thus helped to improve the manuscript enormously.  MS acknowledges supports from NASA/Chandra
grants AR9-0013X and GO0-11063X and NSF grant
AST-0909073.  This research has made use of data obtained from the Chandra Data Archive, and software provided by the Chandra X-ray Center (CXC) in the application package CIAO. It was also based on observations made with ESO Telescopes at the Paranal Observatory under programme ID 073.D-0392(A) and (B), and XMM-Newton, an ESA science mission
with instruments and contributions directly funded by
ESA Member States and NASA.

\end{acknowledgements}

\bibliographystyle{aa} 

\bibliography{M22spec_rev2_corr}

\end{document}